\documentclass[12pt]{article}
\usepackage{graphicx}
\usepackage{setspace}
\usepackage[super,sort&compress,numbers]{natbib}
\usepackage{multirow}
\usepackage{bm}
\usepackage{amsfonts}
\usepackage{amssymb}
\usepackage{amsmath}
\usepackage{enumerate,ulem}
\usepackage{gensymb}
\usepackage{natbib}
\usepackage{mathtools}

\usepackage[colorlinks=true,linkcolor=blue]{hyperref}%
\usepackage{color}

\begin{document} 

\begin{center}
{\bf \large The impact of nanoscale compositional variation\\ on the properties of amorphous alloys}
\vspace{0.5cm}

Ryota Gemma$^{1,}$\footnote{Present address: Department of Materials Science, Tokai University, Japan}, Moritz to Baben$^{2,3,}\footnote{Present address: GTT-Technologies, Kaiserstr. 103, 52134 Herzogenrath, Germany}$, Astrid Pundt$^{4}$,\\ Vassilios Kapaklis$^{2}$ and  Bj\"{o}rgvin Hj\"{o}rvarsson$^{2,}\footnote{Corresponding author: \href{mailto:bjorgvin.hjorvarsson@physics.uu.se}{bjorgvin.hjorvarsson@physics.uu.se}}$

{\it \small
\noindent
$^1$ Physical Sciences and Engineering Division, \\King Abdullah University of Science and Technology (KAUST), Saudi Arabia\\
$^2$ Department of Physics and Astronomy, Uppsala University, Box 516, 751 20 Uppsala, Sweden\\
$^3$ Materials Chemistry, RWTH Aachen University, Kopernikusstr. 10, 52074 Aachen, Germany\\
$^4$ Institute of Applied Materials (IAM-WK), Karlsruhe Institute of Technology KIT, \\76131 Karlsruhe, Germany\\
\vspace{0.5cm}
}

\end{center}

\vspace{0.5cm} 
{\bf The atomic distribution in amorphous FeZr alloys is found to be close to random, nevertheless, the composition can not be viewed as being homogenous at the nm-scale. The spatial variation of the local composition is identified as the root of the unusual magnetic properties in amorphous  Fe$_{1-x}$Zr$_{x}$ alloys. The findings are discussed and generalised with respect to the physical properties of amorphous and crystalline materials. }

\section*{Introduction}
The physical properties of amorphous materials are only partially understood and the relation between local atomic arrangements and emergent physical properties are not fully explored. Crystalline materials are readily characterised with respect to structure and chemical composition using standard scattering techniques, while amorphous materials being disordered, do not offer that possibility. Since the pioneering work on the amorphization of metals\cite{Buckel:1952kt} and alloys\cite{KLEMENT:1960ju}, the field has developed substantially and a large variety of methods are now available for the fabrication of amorphous alloys\cite{Greer_Science_1995}. While the end products are not always identical with respect to physical and chemical properties, it is challenging to identify the underlying reason for the observed differences. The atomic arrangements in amorphous alloys are not well-defined, exhibiting close resemblance to liquid like structures\cite{Angell_Science_1995}, rendering the task of linking their structure to the observed physical properties challenging. 

Amorphous materials are known to exhibit extraordinary mechanical properties \cite{review1, review2, greer_ma_2007, mech_light, Demetriou_NatMat_2011} and some of the observations can be rationalised using computational methods, $\it{e.g.}$  linking shear-resistant structural\cite{zhang2009bulk} to the presence of short-range order\cite{Cheng_ActaMaterialia_Mecahnical_2008}. High-density icosahedral packing of atoms, with a Voronoi coordination polyhedron with index $\langle$0,0,12,0$\rangle$, $i.e.$ all nearest-neighbor pairs are five-fold bonds, exhibit the highest resistance to shear transformations, while the less-ordered and less-densely packed regions are easier to shear\cite{zhang2009bulk}. These results have $\it{e.g.}$ been used to explain the temperature dependence of the elastic limit of Co-B metallic glasses\cite{schnabel2015temperature}. 
\begin{figure}
\includegraphics[width=\columnwidth]{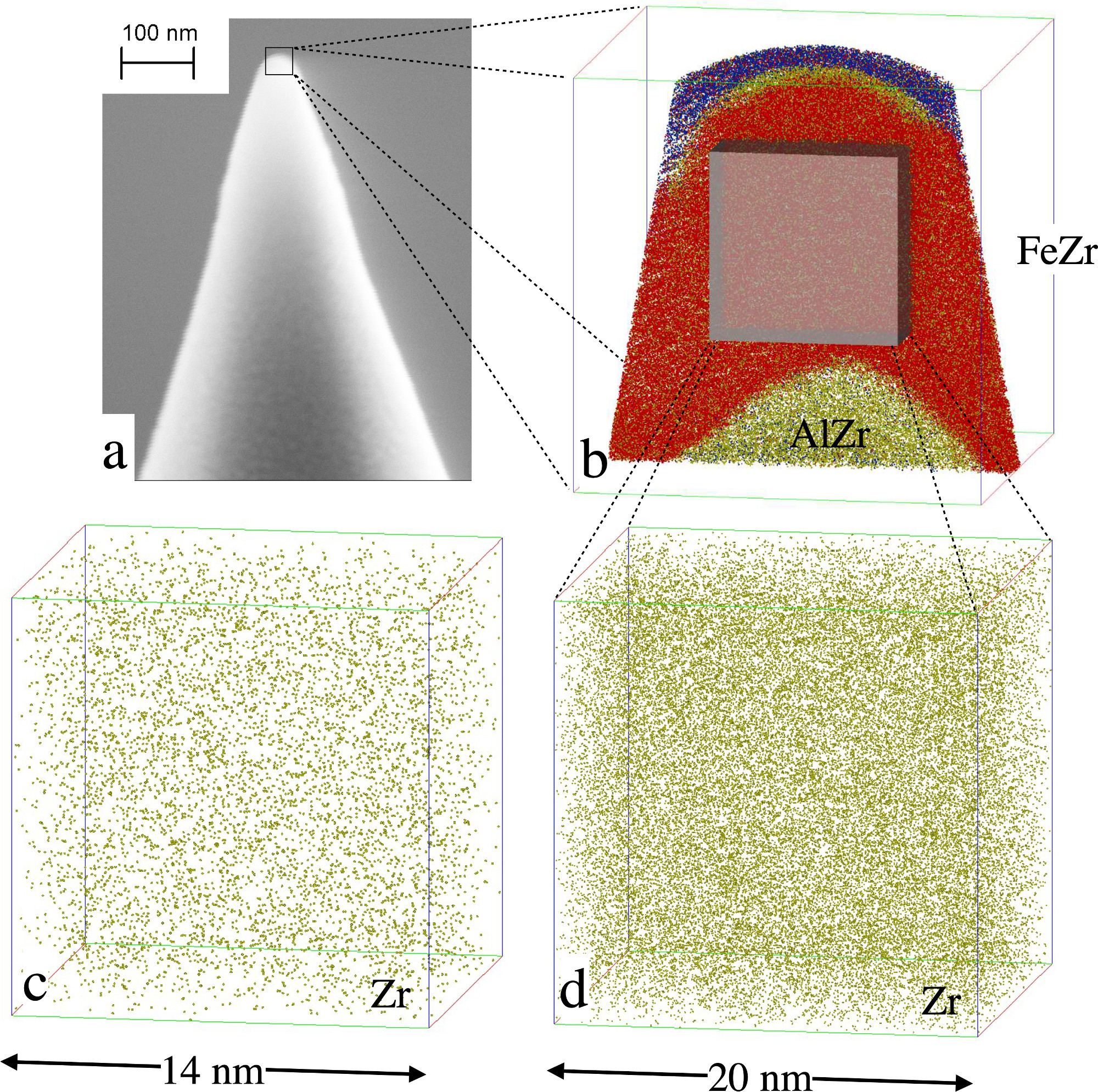}
\caption{{ (a) Illustration of a Si tip covered by an amorphous structure, consisting of three layers. The atomic distribution is illustrated in (b).  At the bottom, the Al$_{75}$Zr$_{25}$  seeding layer (green) is seen, covered by a Fe$_{81}$Zr$_{19}$ (red and green) layer. Topmost the presence of oxygen  (blue) demonstrates the almost complete oxidation of the capping layer (Al$_{75}$Zr$_{25}$ ). In d) the distribution of Zr in Fe$_{81}$Zr$_{19}$ is illustrated and in c) corresponding distribution of Zr is shown  for the Fe$_{91}$Zr$_9$ alloy is displayed.}}
\label{fi:fig_1}
\end{figure} 


Amorphous metals do not only exhibit extraordinary mechanical properties, their magnetic properties are equally unique. For example, metallic glasses can be extremely soft magnetically,  exhibit gigantic magnetic proximity effects\cite{magnus_gig} and have  even been shown to violate Hund's third rule\cite{kapaklis-violation}. 
The spatial variations in concentration and coordination number can be assumed to play a similar role for the magnetic and mechanical properties as $e.g.$ discussed in the analysis of the density, elastic and magnetic properties of CoFeTaB and CoFeTaSi alloys using {\it ab initio} theory\cite{Hostert_2011_JPCM, Hostert_2012_Scripta_Materialia}. This point is also immediately at hand when discussing the proposed magnetic states in amorphous Fe, which depends strongly on the distribution in the alloying element-induced atomic distances in Fe\cite{Krebs1987phaseseparationinFeZr, xiao1987nonuniqueness, bakonyi2012relevance}.  
While the spatial variation in atomic density and coordination number are used for rationalising the mechanical and magnetic properties of amorphous materials, the experimental determination of these is scarce.
The lack of translational and rotational symmetry renders the experimental study of their atomic structure highly challenging: Due to the absence of long range order in amorphous materials conventional diffraction methods yield limited information. However, recent development within the field of nano-beam electron diffraction (in a transmission electron microscope) has enabled direct observation of the local atomic order in a metallic glass \cite{hirata2011direct}. Furthermore, the local atomic structure, including the local configuration numbers could be determined.  But this approach does not  provide information on the local atomic density nor the spatial variation in the  chemical composition. 
Here, we utilize one of the best studied amorphous magnetic alloys\cite{Buschow_FeZr_PhaseDiagram_1981}, Fe$_{1-x}$Zr$_{x}$, to adress  the relation between the distribution of the elements and the observed magnetic properties\cite{martina_critical,martina_reversed,magnus_prox},  using Atom Probe Tomography (APT). We generalise our findings and contribute thereby to the formation of a conceptual base for the understanding of the physical properties of amorphous alloys.

\section*{Results and discussion}

\begin{figure}
\includegraphics[width=\columnwidth]{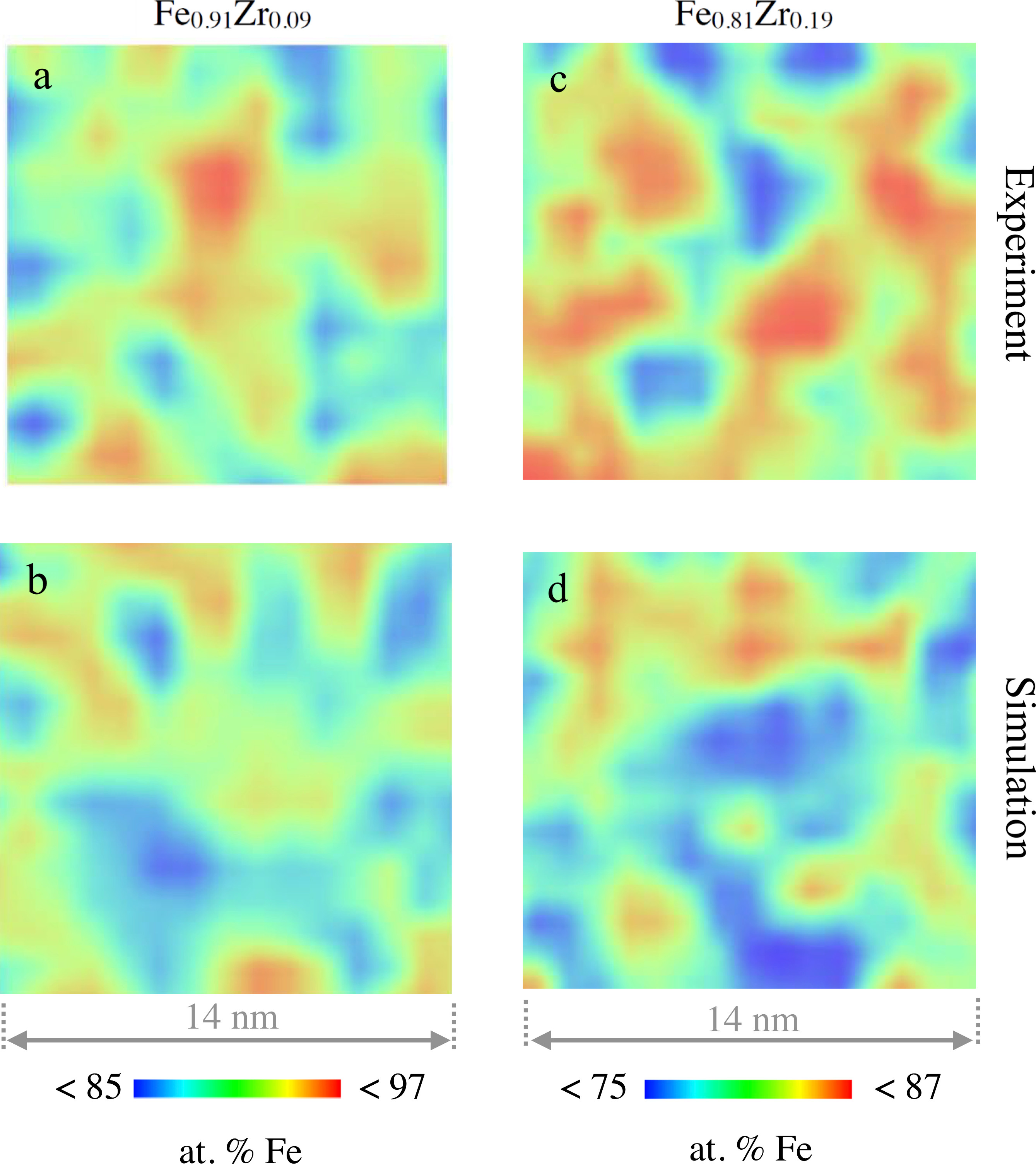}
\caption{{ Measured 2-D Fe contour maps of Fe$_{0.91}$Zr$_{0.09}$ (a) and Fe$_{0.81}$Zr$_{0.19}$  (c). The size of the images is 14 nm x 14 nm. For comparison, simulated 2-D Fe contour maps for a random solution of same composition are included, (b) and (d).}}
\label{fi:fig_2}
\end{figure}

Typical reconstructions of the elemental distribution are shown in Figure~\ref{fi:fig_1}. 
Since the samples were deposited on pre-sharpened Si tips, the interfaces between the layers are curved, reflecting the initial surface geometry as seen in Figure~\ref{fi:fig_1} (a) and (b). The red region in Figure~\ref{fi:fig_1} (b) marks the Fe$_{1-x}$Zr$_{x}$-layers, the yellow regions represent the amorphous Al$_{70}$Zr$_{30}$  buffer and the blue regions marks the partially oxidised  Al$_{70}$Zr$_{30}$  capping layers. The measured Zr distribution in Fe$_{0.81}$Zr$_{0.19}$ and Fe$_{0.91}$Zr$_{0.09}$ are displayed in Figure~\ref{fi:fig_1} (c) and (d), respectively, within which the difference in the Zr-density of the samples is easily seen. When the local concentration of Fe  is displayed in a similar way, the (high-) Fe density hinders any meaningful comparison between the samples. Thus, to illustrate the Fe distributions we need to invoke a different approach: We averaged the Fe concentration across 2 nm thick segments, thin enough to avoid severe blurring of the lateral changes in the composition, while providing statistically significant results, as illustrated in Figure~\ref{fi:fig_2}. Experimental contour maps are shown on the top (Figure~\ref{fi:fig_2} (a) and (c)) while the illustrations at the bottom (Figure~\ref{fi:fig_2} (b) and (d)) displays identical analysis of simulated random distributions of the elements (see Methods). Here it becomes clear that a random distribution does not result in a homogenous concentration of the constituents. A clear variation in the atomic densities is seen on the length scale of few nanometers. 

\begin{figure}
\includegraphics[width=\columnwidth]{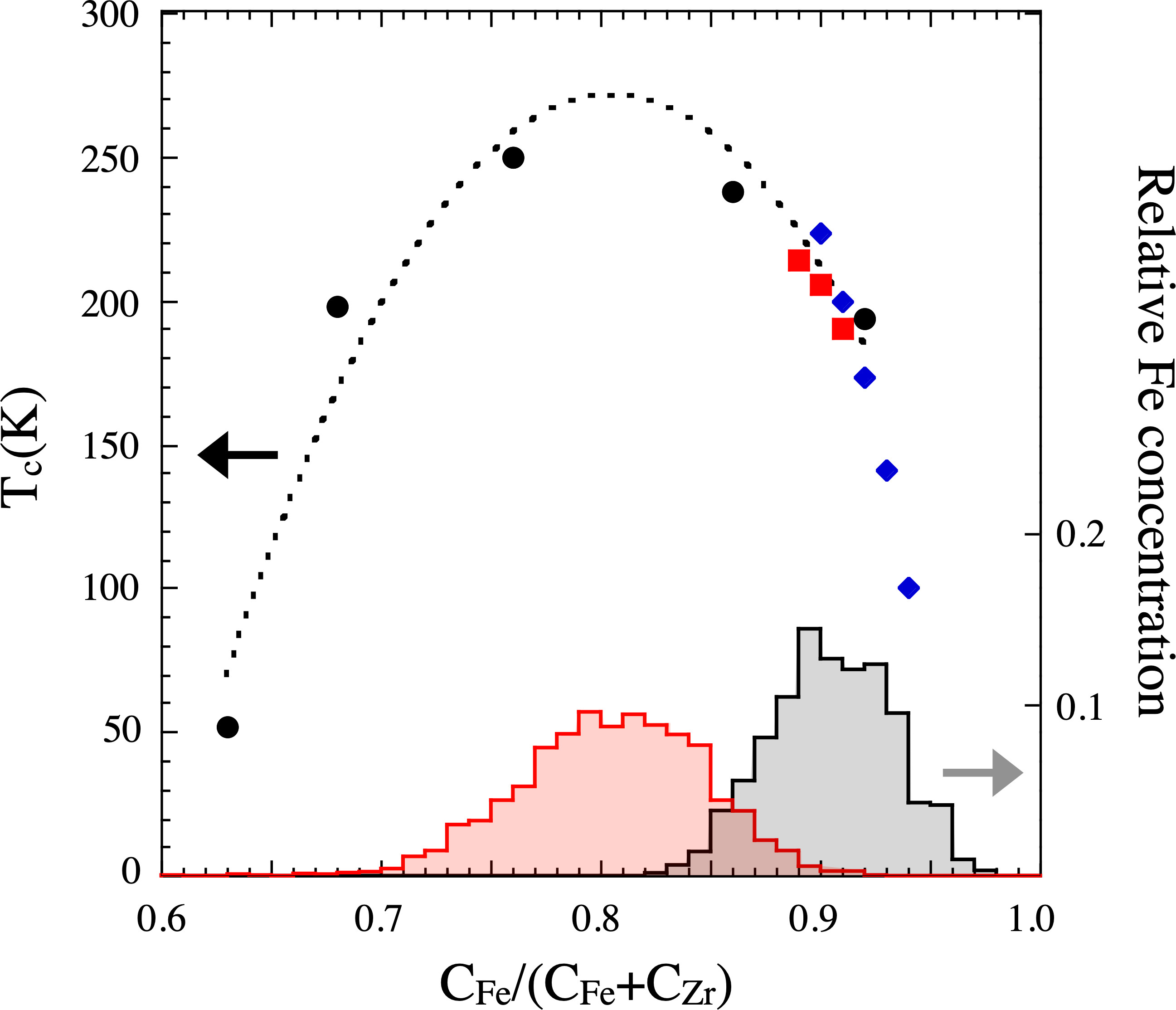}
\caption{{Illustration of the changes in T$_{c}$ (left y-axis) with Fe concentration and the local abundance of Fe in the samples (right y-axis). The black circles (fitted) are from Sharma {\it et al.} \cite{PRBTc}, the blue diamonds are from Read {\it et al.} \cite{JMMMTc} and the red squares are from Korelis {\it et al.} \cite{panos_thesis}.  The shaded area shows the relative abundance of the local Fe concentrations in both the samples (the areas are normalized to unity). Average Fe concentrations of $\gtrsim$0.93 results in crystallisation of the alloy. }}
\label{fi:fig_3}
\end{figure}

The observed length scales in the contour maps of Figure~\ref{fi:fig_2} are all similar. However the distribution in the relative elemental abundance (Fe and Zr) are somewhat different, as $e.g.$ observed in the Fe distribution illustrated in Figure~\ref{fi:fig_3} (right axis). This is possibly reflecting a contribution from a thermodynamic driving force arising from the concentration dependence in the mixing enthalpy of the elements. Let us now consider which effect the spatial variation in concentration can have on the magnetic properties of amorphous Fe$_{1-x}$Zr$_{x}$ alloys. When adding a non magnetic element to a ferromagnetic material, the magnetic ordering temperature (T$_{c}$) typically  decreases monotonically over a wide concentration range. This effect can be viewed as a consequence of decreased magnetic interactions ($\it{J}$) upon dilution as T$_{c} \sim J$ in a homogenous magnetic system. Fe$_{1-x}$Zr$_{x}$ alloys exhibit richer concentration dependence, as illustrated in Figure~\ref{fi:fig_3}, in which a maximum in T$_{c}$ is observed, for an Fe concentration of $\simeq$ 0.8. We can use these results to calculate the strength of the local magnetic exchange interaction based upon the concentration maps depicted  in Figure~\ref{fi:fig_2}. 
To do so we make an ansatz: T$_{c} \sim <J>$, where the brackets denote a weighted average with respect to concentration. Thus the determined T$_{c}$ is assumed to reflect an average exchange coupling dictated by the average concentration within each voxel.

Figure~\ref{fi:fig_3} shows both the concentration dependence of T$_{c}$  (left hand y-axis) as well as the determined distribution of Fe concentrations within the samples (right hand y-axis). Although the variance in the distribution is not negligible, we argue the calculations can be used to map the local coupling strength from the average T$_{c}$ values.
To ease the comparison, we define the local magnetic interaction, $J_i$, in units of temperature. Based on the above assumptions we calculated the local exchange coupling for both the samples and the results are illustrated in Figure~\ref{fi:fig_4}. In these calculations we have used an interpolation and extrapolation for concentrations above 0.93 (see Figure~\ref{fi:fig_3}). This is not expected to change the interpretation in any qualitative way, although we can not exclude changes (errors) in the calculated values of $J_i$. As seen in the figure, $J_i$ is changing dramatically ($\Delta J_i \approx$ 130 K) over short distances in Fe$_{0.91}$Zr$_{0.09}$, forming twined magnetic regions, resembling the contour maps of the elemental concentrations. The magnetic properties can therefore not be viewed as being homogenous, even on the length scale of few nm. The results obtained from the Fe$_{0.81}$Zr$_{0.19}$ sample are illustrated in the right hand part of Figure~\ref{fi:fig_4}. The range in $J_i$ is much smaller ($\Delta J_i \approx$ 46 K) as compared to Fe$_{0.91}$Zr$_{0.09}$. The change in effective exchange coupling with concentration ($\delta J / \delta c$) is therefore a measure of how corrugated the energy landscape will be. These changes in magnetic interactions must be reflected in $\it{e.g.}$ the changes in the spontaneous magnetisation with temperature and we would expect the largest effects to be seen in Fe$_{1-x}$Zr$_{x}$ samples when x~$\lesssim$~0.7 and  x $\gtrsim$ 0.9. 
Let us now test these ideas by comparing the magnetic properties of thin amorphous layers and their single element crystalline counterparts. The ordering temperature of magnetic and structural phase transitions in thin layers are found to scale with the thickness ($n$)\cite{PRL_BH} and can be described as:

 \begin{figure}
\includegraphics[width=\columnwidth]{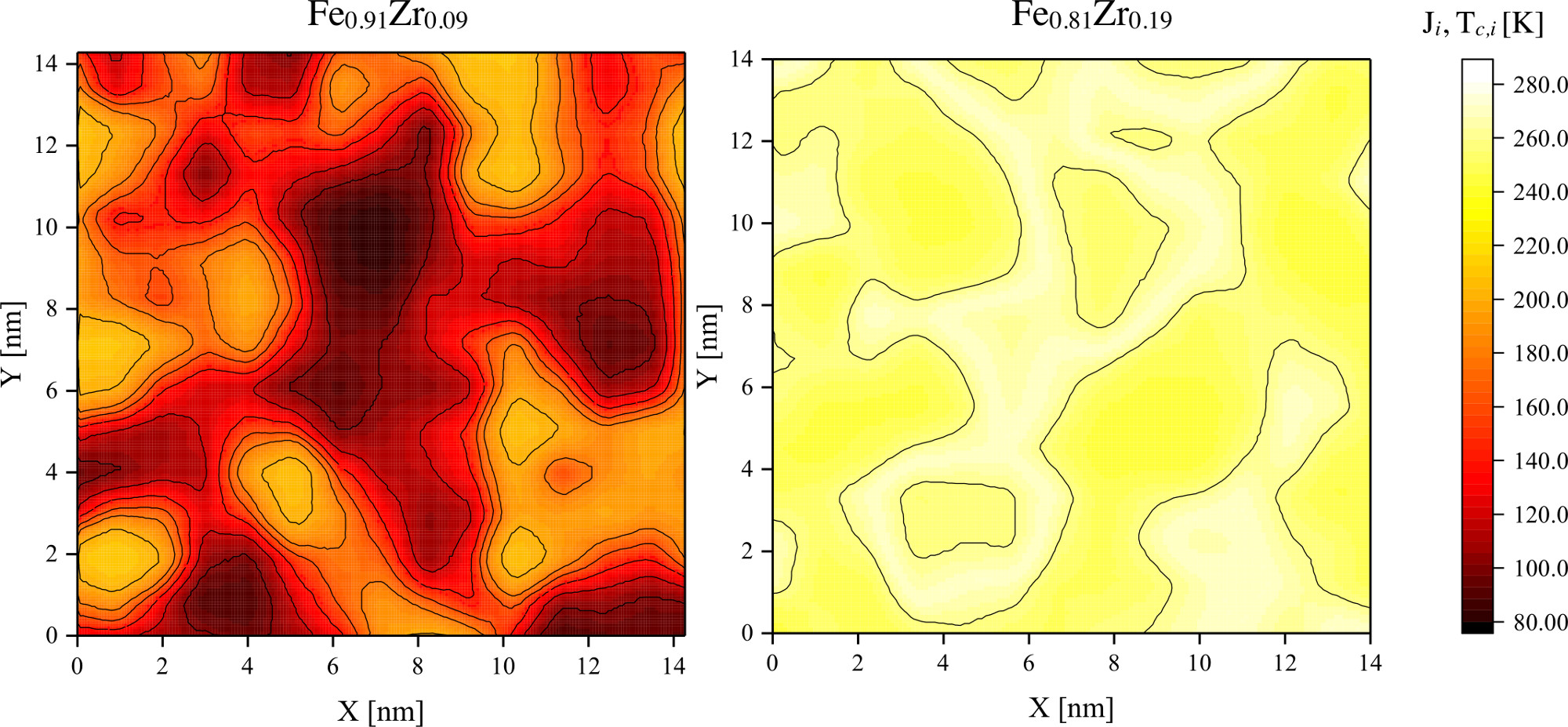}
\caption{{ Changes in the effective coupling strength with concentration in the Fe$_{0.91}$Zr$_{0.09}$ (left) and Fe$_{0.81}$Zr$_{0.19}$ (right) alloys, expressed as T$_{c,i}$. The contour lines in both colormaps depict isolines with an interval  of $\Delta$T$_{c,i}$ = 20 K.} Large difference between these concentrations are inferred, reflecting the change in the effective exchange  coupling with concentration. }
\label{fi:fig_4}
\end{figure}

\begin{figure}
\includegraphics[width=\columnwidth]{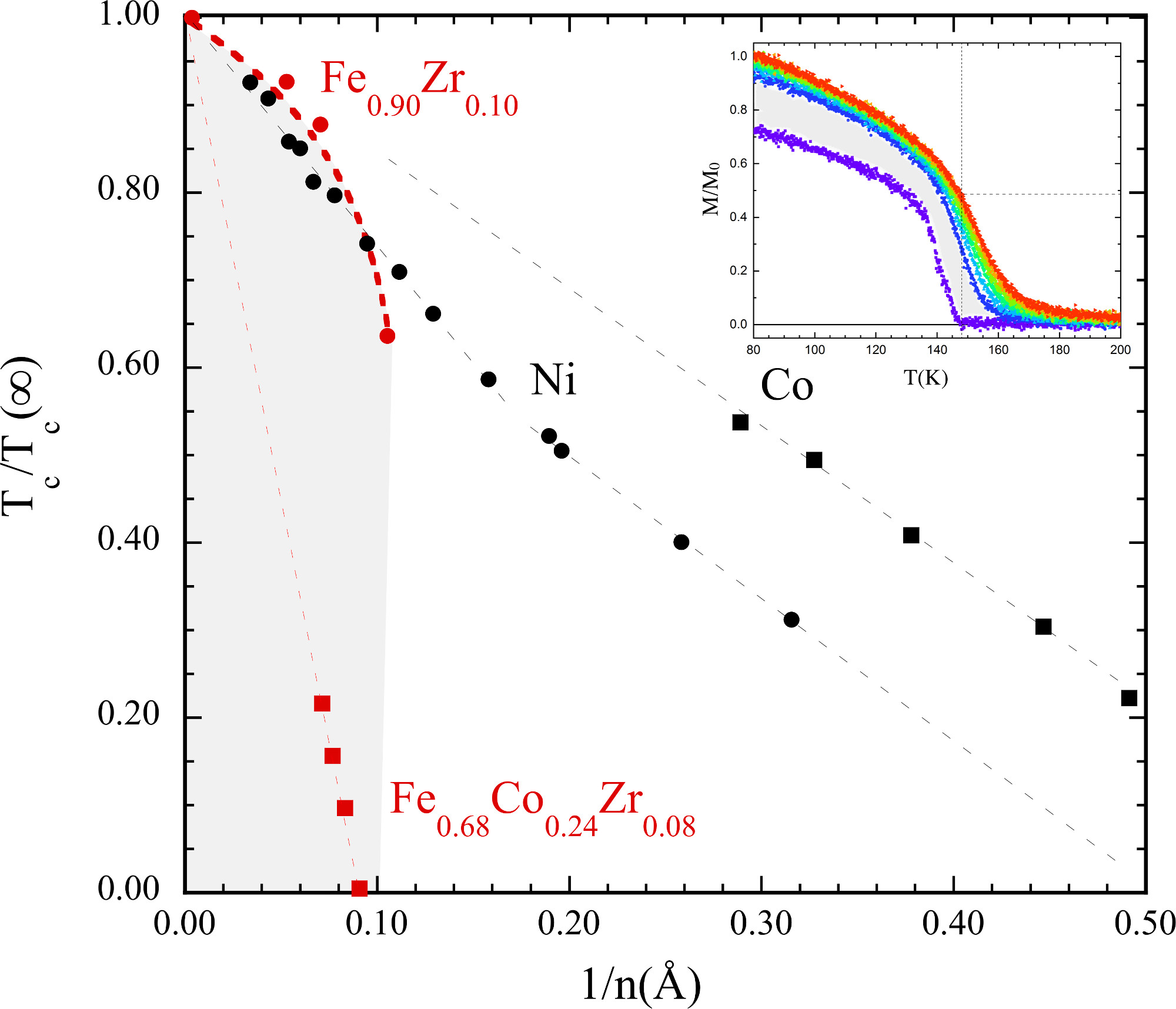}
\caption{{Illustration of the changes in T$_c$(n)/T$_C$($\infty$) with inverse thickness of crystalline and amorphous layers. The Ni and Co data are adapted from Huang {\it et al.} \cite{PRB_Willis}, the FeCoZr from Ahlberg {\it et al.} \cite{PRB_martina_fecozr} and Fe$_{0.90}$Zr$_{0.10}$ from reference Korelis {\it et al.} \cite{FeZr_finite_1}. The inset illustrates the the temperature and field dependence of the magnetisation in a 1.5 nm Fe$_{0.89}$Zr$_{0.11}$ at fields between 0 and 6 mT adapted from Liebig {\it et al.} \cite{PRB_liebig}. }}
\label{fi:fig_5}
\end{figure}

\begin{center}
 T$_{c} (\it{n})$/T$_{c} (\infty)$ = $(1- \frac{1+2\Delta n}{n})^\lambda$,
 \end{center}
 
 \noindent where $\Delta n$ is the extension of a ``dead" layer at each interface, $\lambda$  is an exponent and T$_{c} (\infty)$ is the ordering temperature of bulk (infinitely large) sample. Typical results obtained from crystalline single element layers and alloys of amorphous materials are illustrated in Figure~\ref{fi:fig_5}. The results obtained from crystalline Co and Ni on Cu\cite{PRB_Willis}, as well as Fe$_{0.68}$Co$_{0.24}$Zr$_{0.08}$\cite{PRB_martina_fecozr} layers are reasonably linearised over a wide range in this representation (1/$n$). The changes obtained from  Fe$_{0.90}$Zr$_{0.10}$\cite{FeZr_finite_1} layers, exhibit completely different behaviour, with $\lambda =$ 0.16$\pm$0.04 as compared to $\lambda \simeq$ 1 for the other layers. This is not surprising when considering the extreme variation of the effective coupling strength within the Fe$_{0.91}$Zr$_{0.09}$ samples. Extrapolating the thickness dependence of T$_{c}$ for the Fe$_{0.68}$Co$_{0.24}$Zr$_{0.08}$ layers\cite{PRB_martina_fecozr}, results in a T$_{c}$=1025$\pm$7 K which is an order of magnitude higher than that of Fe$_{0.90}$Zr$_{0.10}$. Hence although the concentration dependence of T$_{c}$ is not known, we can safely conclude that  $[\delta J / \delta c]/J$ is at least an order of magnitude larger in Fe$_{0.90}$Zr$_{0.10}$ as compared to Fe$_{0.68}$Co$_{0.24}$Zr$_{0.08}$. This observation provides the basis for the obtained differences and consequently  Fe$_{0.90}$Zr$_{0.10}$ can only been regarded as magnetically continuous well below its ordering temperature. 

The extension of the ``dead" layers, $\Delta n$, is significantly different in crystalline and amorphous samples as seen in Figure~\ref{fi:fig_5}. While crystalline single-element samples typically exhibit a ferromagnetic behaviour to the monolayer limit, amorphous layers loose their spontaneous magnetisation at thicknesses which are almost an order of magnitude larger. The large $\Delta n$ in amorphous alloys is readily rationalised when considering the changes in the effective exchange coupling, reflected in the variation of $J$ within the samples (see Figure ~\ref{fi:fig_4}). Above the apparent T$_{c}$, the amorphous layers will not be paramagnetic: There will be regions with substantial moments, albeit fluctuating, and thereby not contributing to the spontaneous magnetisation. These are separated by sections with a weaker exchange coupling, effectively decoupling the intrinsically ferromagnetic regions. This interpretation is confirmed by the field dependence of the magnetisation of  Fe$_{0.90}$Zr$_{0.10}$, which resembles a super-paramagnetic like behaviour well above the as determined T$_{c}$\cite{FeZr_finite_1, PRB_martina_fecozr, PRB_liebig}. The effect is illustrated in the inset in Figure~\ref{fi:fig_5}, in which a field of 1 mT is seen to induce a moment which is approximately one half of what is obtained at 80 K. The range of the magnetic correlation in these layers, was estimated to be of the order of 100 nm\cite{FeZr_finite_1} at T= T$_{c}$+20 K, which is substantially larger than the length scales of the compositional contours observed here. Thus, well above the ordering temperature there are large regions within which the variations in $J$ are partially suppressed by magnetic proximity effects\cite{magnus_gig}. Furthermore, the large magnetic susceptibility observed in a wide temperature range below T$_{c}$, reflects at least partially the distribution in  $J_i$ (T$_{c,i}$)\cite{FeZr_finite_1, PRB_martina_fecozr}. Finally, when the thickness of the amorphous layers is smaller or equal to  2$\Delta n$, a  superparamagnetic behaviour is observed at 5 K.\cite{FeZr_finite_1} Similar effects are observed in Fe$_{0.68}$Co$_{0.24}$Zr$_{0.08}$ layers\cite{PRB_martina_fecozr}.
The results presented here provide therefore a base for the understanding of the ordering and phase transitions in amorphous alloys, including finite size effects on magnetic ordering. 

\section*{Conclusions}
The randomness in the local chemical composition has a large impact on the magnetic properties of amorphous alloys. Its effect is clearly seen in both finite size scaling of the ordering temperature as well as the extension of interface regions in $e.g.$ Fe$_{1-x}$Zr$_{x}$ amorphous alloys. 
The extraordinary mechanical properties of amorphous alloys\cite{Sheng_Nature_2006, Cheng_ActaMaterialia_Mecahnical_2008} can be argued to stem from the same roots. The analogy to magnetic properties is straight forward: Replacing the magnetic interactions with chemical binding, results in spatial variation of atomic interactions and thereby changes in local mechanical properties. 
Recently, it was noted that atomic arrangements and the related probability distributions for particle displacements can be correlated with string-like excitations. These have a significant impact on the structural relaxation, atomic rearrangement and mechanical properties of metallic glasses\cite{Samwer_SciAdv_2017}. Thus, having access to direct information on the atomic arrangements, such as obtained when using APT, can therefore shed light on a series of open questions concerning the physical properties of amorphous alloys\cite{Berthier_glass_review_2011}. 
We also note the lack of a theoretical framework for both the effect of non-homogenous interactions and its influence on the emergent magnetic order in finite size systems. Finally, to implement realistic descriptions of amorphous alloys we need to recognise that random compositions are intrinsically inhomogeneous in nature. \\

\section*{Methods}

Amorphous FeZr thin films have been deposited by DC magnetron sputtering from elemental targets at room temperature. The base pressure was below $5\cdot 10^{-10}$ mbar and the (purified) Ar pressure during growth was $4\cdot 10^{-3}$ mbar. Since FeZr thin films on Si substrates grow partly crystalline at room temperature \cite{korelis2010highly} an amorphous AlZr seed layer was deposited from an Al$_{75}$Zr$_{25}$ compound target. The same target was used to deposit a capping layer to avoid oxidation of the magnetic layers. For APT, layers were deposited directly on pre-sharpened Si micro-tips with two different Fe and Zr target power ratios, resulting in compositions of Fe$_{91}$Zr$_9$ and Fe$_{81}$Zr$_{19}$. The chemical compositions were confirmed by energy dispersive X-rays as well as atom probe analysis.

APT analyses were carried out on LEAP 4000 XHR (Cameca) in laser pulsing mode using a laser wavelength of 355 nm, a laser energy of 70 pJ, a pulse repetition rate of 200 kHz, with a detection rate of 0.003 ions per pulse.  Three different samples were successfully analyzed. These contained  2.3 x 10$^5$ atoms in a volume of 20 nm x 20 nm x 20 nm and 7.4 x 10$^4$ atoms in a volume of 14 nm x 14 nm x 14 nm. These were analyzed by two different slices each and in more than six slice volumes, confirming the presented result. The sample temperature was set to 70 K in these measurements. As the pulse method always removes the particular uppermost surface atoms, the depth resolution of this APT analysis is one atomic layer. The lateral resolution within the layer is about 0.5 nm \cite{kelly2007atom}. 

The 3-D reconstruction of the ion positions was performed using IVAS 3.6.6 (Cameca). The initial radius of curvature r$_0$ and the specimen's shank half angle $\theta$ were determined by scanning electron microscope (SEM) (Helios, FEI) before the analysis and later applied in the reconstruction process. In both the amorphous alloys typical values of $r_0$ and $\theta$ were found to be $r_0$ = 30 nm and $\theta$ = 16$\,^{\circ}$, respectively. After the reconstruction, the 1st FeZr layer was chosen for evaluation of the chemical homogeneity by studying the concentration histogram, by using the 'cluster search' and the 'concentration mapping' provided by IVAS 3.6.6. 
In the voltage curve and detection rate curve of analysis, no burst was detected in the FeZr layer. Nevertheless, we cropped a volume of FeZr layer away from AlZr/FeZr interfaces for data evaluation to avoid possible impact of bursts at the proximity of interfaces. Impacts by orientation-dependent differences in resolution on the results could be excluded by using the cluster search.
To allow for undoubtedly atomic classification, the signal at m/e = 27 is removed in the data evaluation processes because of the mass overlap of Fe and Al ($^{54}$Fe$^{2+}$ and $^{27}$Al$^{+}$) in the seeding and capping layers. This approach was not implemented in the analyses of the Fe-Zr layers.

\begin{figure}
\centering
\includegraphics[width=\columnwidth]{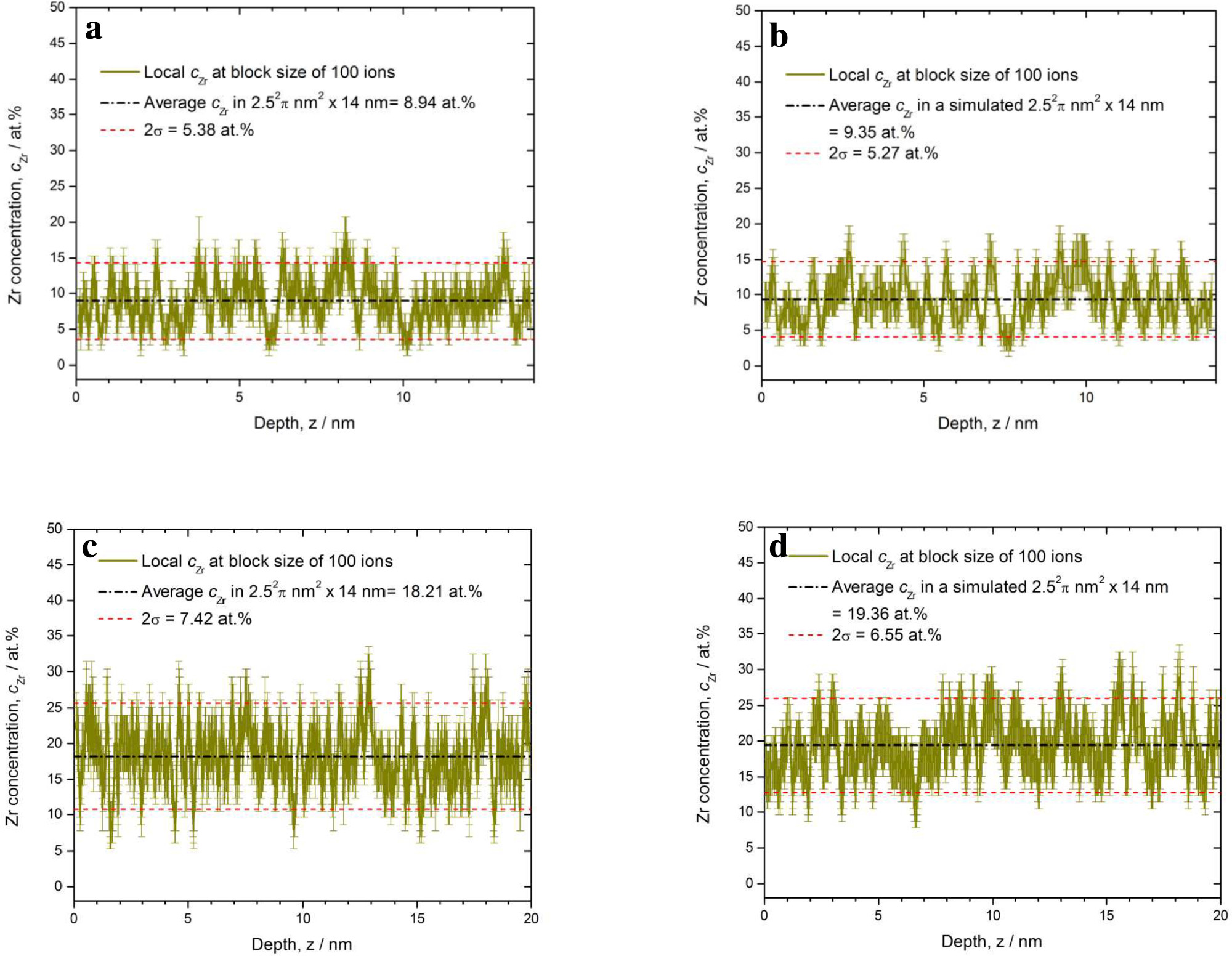}
\caption{{(a) Measured concentration depth profile of Fe$_{0.91}$Zr$_{0.09}$ alloy. (b) Simulated depth profile from a simulated volume of Fe$_{0.91}$Zr$_{0.09}$. (c) Measured concentration depth profile of the Fe$_{0.81}$Zr$_{0.19}$ alloy (d) Simulated depth profile of the Fe$_{0.81}$Zr$_{0.19}$ alloy. The concentration was obtained by averaging over 100 atoms. No differences are observed between the measured and the simulated alloy in a concentration depth profile of this length scale.}}
\label{fi:concprof}
\end{figure} 

\begin{figure} 
\includegraphics[width=\columnwidth]{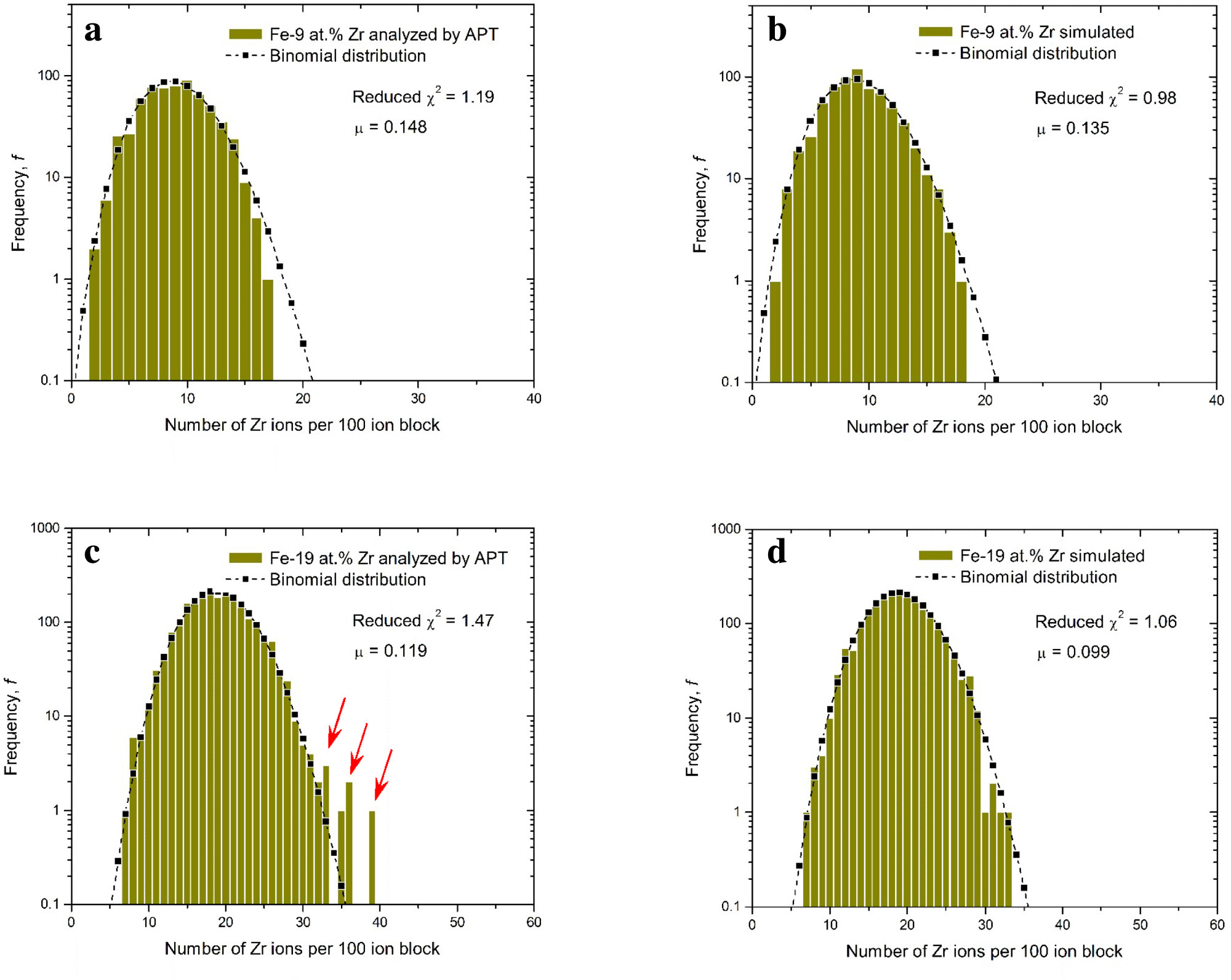}
\caption{{ Measured Zr concentration histogram in a cubic volume of Fe$_{0.91}$Zr$_{0.09}$ (a) and Fe$_{0.81}$Zr$_{0.19}$ (c). For comparison, the simulated Zr concentration histograms for a random solution of similar size and the respective compositions are included in (b) and (d). The block size is 100 ions.}}
\label{fi:histograms}
\end{figure}

A random FeZr alloy with the same volume and nominal composition was simulated and investigated for comparison, by using Region of Interest (ROI) simulation tool of the same software. The simulation of the volume assumed a bcc lattice. We, therefore, included a 0.5 nm smearing of the data to better mimic the amorphous alloys. The detection efficiency was set to 0.36, which is a typical value for the LEAP 4000 with reflectron. The detection efficiency has pure geometrical reasons and is therefore assumed to be insensitive to the detected elements  \cite{kelly2007atom}. The chosen density $\rho$  in atoms per nm$^{3}$ was adjusted to the best match value in a respective volume to that of the measured counterpart. The density is $\rho$ = 77.82 atoms nm$^{-3}$ for Fe$_{91}$Zr$_9$, $\rho$ = 79.86  atoms  nm$^{-3}$ for Fe$_{81}$Zr$_{19}$, respectively.
Typical depth profiles are shown in Figure~\ref{fi:concprof} for a) Fe$_{0.91}$Zr$_{0.09}$  and c) Fe$_{0.81}$Zr$_{0.19}$ samples. The Zr depth profiles shown in Figure~\ref{fi:concprof} a) provide the matching average concentration of 9 at\% Zr for the alloy, as given by the dashed black line. Some local concentration values exceed the doubled standard deviation (2 $\sigma$-value, marked with the red dotted lines) of the average Zr concentration. This is also observed for the simulated alloys illustrate in Figure~\ref{fi:concprof} b) and Figure~\ref{fi:concprof} d).

Figure~\ref{fi:histograms} shows a typical frequency distribution analysis of the measured (left side, a) and c)) and the simulated random (right side, b) and d)) alloy yielding the same average composition. Each block contains 100 atoms. The majority of blocks in the frequency distributions follow the binomial distribution, given by the black dashed envelope. For the measured 19 at.\% Zr alloy shown in Fig.3 c, Zr-rich regions are observed (marked with red arrows) that exceed the binomial envelope and that are not visible for the simulated random alloy. This observation is consistent with a slight thermodynamically driven composition variations in the alloy.

\section*{Acknowledgments}
This work was funded by the Swedish Research Council (VR) and the Foundation of Strategic Research (SSF). Work undertaken at the KAUST was supported by the King Abdullah University of Science and Technology (KAUST). A.P. was supported by the Deutsche Forschungsgemeinschaft in the Heisenberg progam via PU131/9-2. AP and RG thank Dr. T. Boll (KNMF at KIT) for kind discussion. 

\footnotesize{

\section*{Author contributions}
B.H., M. B. and A.P. designed the experiment. M.B., A.P. and R.G. carried out the APT sample preparation. R.G. carried out the APT measurements. R.G. and A.P. performed the APT data analysis. B.H., M. B., A.P.,R.G. and V.K.  contributed to the interpretation of the results and the writing of the manuscript.}

\clearpage

\bibliographystyle{natphys}

\end{document}